# A general method for calibration of active scanning thermal probes

Alexander Tselev[*]

*CICECO – Aveiro Institute of Materials and Department of Physics,*

*University of Aveiro, 3810-193 Aveiro, Portugal*

**Abstract.**

Scanning Thermal Microscopy (SThM) is a scanning probe technique aimed at quantitative characterization of local thermal properties at the length scale down to tens of nanometers. With many probe designs and approaches to interpretation of probe responses, there is a need for a universal framework, which would allow probe calibration and comparison of probe performance. Here, we have developed a calibration framework based on an abstracted, formal, probe model for active SThM probes. The calibration can be accomplished through measurements with two or three calibration samples. Requirements to calibration samples are described with examples of structures of suitable samples identified in published literature. A link to a published experimental work indirectly verifying the proposed procedure is provided. The calibration does not require knowledge of internal probe properties and yields a small and universal set of parameters that can be used to quantify thermal resistance presented to the probe by samples as well as to characterize active-mode SThM probes of any type and at any measurement frequency. We have illustrated how the probe calibration parameters can be used to guide probe design. We have also analyzed when the calibration approach can be used directly to measure thermal conductivity of unknown samples.

## 1. Introduction

Thermal management and control of heat generation, dissipation, and transfer at the micrometer and submicrometer are factors of critical importance for correct and reliable operation of modern and future electronic devices and sensors,[1] thermoelectric energy conversion,[2] phase change memory systems,[3] and other systems, where functionality is based on submicrometer-scale structures. On the other side, details of the heat transfer through nanometer scale extended defects in materials, such as grain boundaries in polycrystals, are of importance for engineering and application of functional materials, also in a bulk form.[4] Understanding, engineering, and control of nanoscale thermal processes demand quantitative characterization of temperature distributions as well as associated material properties at the nanometer scale. Scanning probe microscopy (SPM) has established itself as a suite of versatile techniques based on local probing and mapping of materials at the nanoscale via measurements of material-probe interaction with probes sensitive to specific functional material properties and parameters of the material state.[5, 6] The SPM methods developed for thermal characterization are generally referred to as Scanning Thermal Microscopy (SThM) and include probing techniques based on different physical principles.[7-12] An advantage of SThM over other methods is a higher spatial resolution with an ability to tackle material characterization at length scales down to 100 nm and below, which is above capabilities of, e.g., optical methods, such as thermoreflectance and optical spectroscopy techniques,[4, 13] where spatial resolution is light-diffraction-limited. In turn, in SThM, the spatial resolution is defined by the probe size and measurement sensitivity. SThM in different embodiments was applied to mapping temperature distributions and thermal conductivity in various nanoscale objects, including active devices,[14-19] one- and two-dimensional (2D) materials,[20-31] thin films,[32, 33] resistive switching structures,[34, 35] as well as effect of doping and processing in inhomogeneous materials,[36] revealing a wealth of thermal processes and properties at the nanoscale.

---

[*] E-mail: atselev@ua.pt

SThM probing assumes detection of nanoscale probe temperature, which is a function of sample local temperature and thermal conductivity. The detection can be realized using different approaches. The most widely used to date exploits thermoresistance, that is, dependence of electrical resistance on temperature, when probes are equipped with thermistors.[7, 33, 37-43] Thermoresistive probes with thermistors made of metals[41, 44, 45] and doped silicon[37, 39] were implemented. A related approach with a sensor made as Schottky diode was reported in Ref. [46]. Another widely used temperature sensing principle is thermoelectric, when a nanoscale sensor is realized as a thermocouple, whose electromotive force is a function of temperature.[47-52] Mechanical temperature detection was also demonstrated with bimorph cantilever probes with composite cantilevers made of layers or legs of materials with different thermal expansion coefficients.[53-56] Temperature-dependent bending of such cantilevers is detected with a standard optical lever system of an atomic force microscope (AFM). As another possibility, a fluorescent particle glued at the end of an AFM probe tip were used as a temperature sensor in Ref. [57] due to a temperature-dependent fluorescent light intensity.

SThM can be performed in two modes: active and passive. In passive mode, the probe is used as a temperature sensor only. In this mode, the temperature field of a heated sample is mapped. In turn, in the active mode, the probe is the temperature sensor and simultaneously, it is a localized scanning heater of a sample.

The active-mode SThM is used for probing and mapping thermal conductivity through detection of variations in heat flow into a sample from a heated probe; the detection is based on sensing variations of the probe temperature near the probe-sample contact point. To date, a quite broad spectrum of probe designs for active-mode SThM with different approaches both for probe heating and for temperature sensing were implemented or demonstrated. A probe can be heated with an electric current[37, 42, 43, 58] or with a laser beam,[52, 54] the temperature changes can be detected via changes of electrical resistance,[37, 42, 43] thermoelectric voltage of a thermocouple junction,[52, 58] or mechanical bending of the cantilever probe.[55] For interpretation of the measurement and to achieve quantification of sample thermal properties, generally, performance of probes is analyzed with the help of analytical or numerical models capturing internal processes inside probes with models developed specifically for each probe design and configuration.[20, 42, 58-65] With the variety of operational principles and complexity of the probe models, it becomes evident that it is highly desirable to have a broadly applicable framework that would allow comparing performance of different probes with the use of a relatively small set of parameters as well as directing their design and optimization. Similarly, a unified approach to probe calibration for quantitative measurements of thermal conductivity would be highly beneficial.

In this paper, we have developed such a framework. The analytical approach employed here to this end is routinely used in electric engineering, especially in high-frequency and microwave measurements, and mathematically based on a matrix formalism. Due to the mathematical identity of the electric current and heat conduction equations, the matrix formalism of electrical circuits can be analogously applied for analysis of heat transport in compound systems.[66, 67] An SThM probe interacting with a sample is just an example of such a system, and matrix-based (quadrupole) models have been developed, in particular, for nanofabricated thermistor-based active SThM probes.[68] The well-established power of the matrix formalism in the description and characterization of electrical circuits in that the circuit representation with matrixes removes both the need to solve circuit or Maxwell equation for internal parts of the circuits as well as the need-to-know details about their internal electrical properties, which drastically simplifies circuit design and measurement tasks.[69, 70] Therefore, the potential power of this approach when applied to the SThM, especially from the user perspective, is that it eliminates the need to have detailed information about internal state and properties of an SThM probe. The formalism, relying on abstracted rather than explicit, operational-principle-specific probe models, can be universally applied independently of the principle of probe temperature measurement, for example, resistive or thermoelectric, even when explicit models are unavailable. It is applicable to stationary measurements as well as measurements with harmonically oscillating stimuli. With the use of this approach, probes can be characterized with a few parameters obtained through measurements of standard, calibration, samples. The parameters allow comparison of probe performance and straightforward calculation of sample-related quantities from measurement data, which can open the door to standardization of probe specifications. Such a standardization will make it possible to predict and compare performance of probes of different types just based on their specifications, which is currently impossible.

While the calibration approach developed here can be applied to measurement both in air and in vacuum, we are concerned in this work only with measurements in vacuum because of a large measurement uncertainty from contributions into heat conduction through air and water meniscus at the tip-sample contact in ambient. According to estimates made in Ref. [71], about 30% of heat generated at the active nanomachined SThM probe is dissipated via air. In Ref. [72], it was shown that in ambient, the sensitivity of the SThM probe in passive mode depends on the heated region size due to the parasitic through-air heat conduction between the probe and the sample, which makes it impossible to perform quantitative thermal measurements in air. However, this effect disappears in vacuum as demonstrated in Refs. [22] and [50].

The outline of the rest of the paper is as follows. In Section 2.1, we reiterate the principle of the active SThM probing and outline the probe calibration approach. In Section 2.2, we introduce the matrix formalism for the SThM probing and derive probe calibration expressions. The calibration expressions are verified with the help of numerical modeling in Sections 2.3 and 2.4. In section 2.5, possible ways to practical implementation of the calibration are discussed with examples of suitable samples from literature. Finally, in Section 2.6, we show as an example how the calibration parameters can be used to compare performance of different probes and to assess probe sensitivity.

## 2. Results and discussion
### 2.1 Principle of SThM in the active mode and approach to probe calibration

The schematic in Figure 1a illustrates the principle of the active-mode SThM probing. A heat power $Q$ is supplied to the probe, normally in the vicinity of its apex, which is in contact with a sample. The supplied heat leaves the probe into the probe base (heat flux $Q_\text{p}$ in Figure 1a), as well as into the sample through the sensing tip and tip-sample boundary (heat flux $Q_\text{s}$). The heat flux $Q_\text{s}$ depends on the thermal conductance of the sample material in contact with the probe, and the temperature along or near the sensing tip, $T_\text{p}$, is monitored providing the information about the sample property.

Figure 1b displays a widely used equivalent lumped-element thermal circuit of an SThM probe operated in the active mode in contact with a sample. The circuit represent the two parallel channels of heat conduction, through which the heat power $Q$ supplied to the probe leaves the probe, with $Q = Q_\text{p} + Q_\text{s}$. The thermal resistance of the probe structure between the heated spot and probe base (Figure 1a), which we will call here *lever*, is represented by the thermal resistance $R_l^{th}$. The second channel, into the sample, is represented by the thermal resistance $R_c^{th}$, i.e, the probe-sample contact structure. This circuit is sufficient for measurement in vacuum. In air, heat transfer through air can be accounted for with additional branches parallel to the lever as well as to the contact structure. In this paper, as mentioned in Introduction, only measurements in vacuum are of interest even though some of the results below can be generalized to measurements in air.

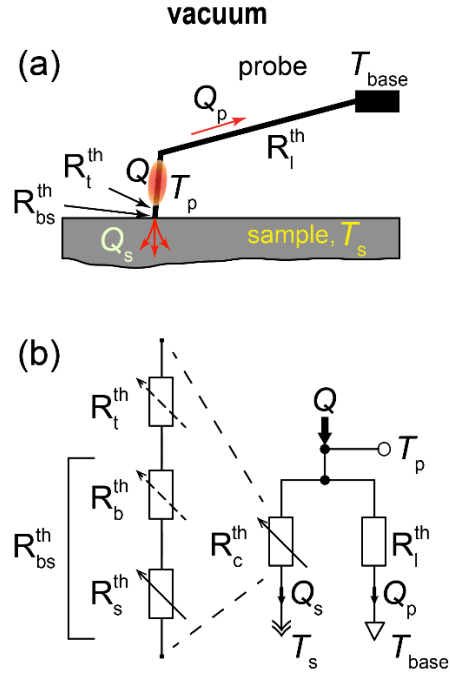

**Figure 1.** (a) Schematic illustrating the active-mode SThM probing and (b) an equivalent lumped-elements thermal circuit of an SThM probe in contact with a sample in the active mode. Generally, temperatures of sample, $T_s$, and base, $T_{base}$, can be different. See text for notations.

In the probe-sample contact structure, at least three thermally resistive components should be distinguished: the part of the probe tip between the heated spot and the probe-sample boundary with a resistance $R_t^{th}$, the probe-sample boundary resistance, $R_b^{th}$, and the spreading resistance of the sample, $R_s^{th}$. As depicted in Figure 1b, they form an in-series connection. The information about sample thermal conductivity is contained in $R_s^{th}$, and its determination requires knowledge of $Q$, $R_l^{th}$, $R_t^{th}$, $R_b^{th}$, as well as temperatures of the probe base, $T_{base}$, of sample, $T_s$, and of a point along the probe tip, $T_p$, which generally may or may not coincide with the heated spot on the probe. During measurements, $Q$ is directly monitored, temperatures $T_{base}$ and $T_s$ are anchored to the temperature of the environment or measured directly with suitable thermometers, while $T_p$ is a temperature of the submicrometer-to-micrometer-large part of the probe near its apex and its determination is a major challenge, which is addressed differently depending on the probe type. Values of $R_l^{th}$ and $R_t^{th}$ are properties of the probe and, in principle, can be obtained individually for each probe via dedicated measurements and/or modeling. In practice, $R_t^{th}$ is difficult to determine experimentally because it is embedded into the sum $R_t^{th} + R_b^{th}$ with $R_b^{th}$ being dependent on the sample and probe-sample contact conditions. Furthermore, $R_b^{th}$ and $R_s^{th}$ are functions of the tip-sample contact size, which spans in SThM from a few tens to a few hundreds of nanometers, and which needs to be determined via dedicated measurements as well.

While the lumped-element thermal circuit in Figure 1b is simple to solve, in practice, in all types of probes, heating and sensing probe temperature $T_p$, take place over a size larger or much larger than the probe-contact size, that is, the size of the probed volume, and heating and temperature sensing functions may or may not overlap in a probe. This means that both probe heating and temperature sensing should be regarded as distributed over some part(s) of the probe with generally different sizes of heated and temperature sensing parts. Since all types of probes are distributed systems, their properties can be captured only approximately with the help of simple lumped elements. At the same time, full-scale models of probes are quite complex with many parameters, which cannot be accurately determined. We note, however, that in SThM, a probe serves as an interface between a signal conditioning/measurement system and a sample, and detailed knowledge of internal properties of this interface is unnecessary. A very similar situation exists in characterization of electrical low- and high-frequency circuits, where connectors, probes, and cables between the circuits and test and measurement devices are all distributed systems and are inevitable interfaces in such equipment as LCR-meters and microwave network analyzers. The exact properties of the interfaces are generally of little or no interest, and their presence and properties are accounted for with a set of calibration measurements, which are sufficient to characterize an interface for a measurement task. Calibration consists of measurement of a small number of well-known, calibration, samples, which may include "open-", "short-" circuit conditions for electrical circuits and simple "loads", such as a 50 Ohm resistor, in place of a tested circuit. During measurements of actual tested devices, calculations are performed by the measurement equipment "on-fly" to "de-embed" the devices from interfaces with the use of the interface parameters known due calibration. The mathematical machinery used to account for the parasitic parameters of interfaces is based on the matrix formalism.[70] Furthermore, the interface parts are characterized by their respective matrix parameters. The ideas behind the calibration techniques used in characterization of electrical networks can be applied to the SThM with thermal probes, which are linear systems in respect to temperature and heat flows.

## 2.2. Matrix formalism and calibration relations for an active SThM probe

With all interface components and tested devices/samples being linear systems, the mathematical formalism treats all components as linear two-port networks (circuits) with two terminals per port (Figure 2a). The two-port networks are considered as "black boxes" described by 2×2 matrixes that linearly relate two input and two output parameters of a network. In electrical circuits, these parameters are currents and voltages. Correspondingly, in thermal circuits, the parameters are heat fluxes and temperatures. Specifically, a port of a heat-conducting network can be described by temperature, $T_i$, and heat flux, $Q_i$, at the port, with $i$ =1 or 2 for input and output ports, respectively (Figure 2a). Depending on the choice of two (out of in total four) variables as independent, different matrices can be used. Here, we will employ admittance matrices $[\alpha_{ij}]$ defined with the following equation:

$$\begin{pmatrix} Q_1 \\ Q_2 \end{pmatrix} = \begin{bmatrix} \alpha_{11} & \alpha_{12} \\ \alpha_{21} & \alpha_{22} \end{bmatrix} \begin{pmatrix} T_1 \\ T_2 \end{pmatrix}, \qquad (1)$$

and transmission matrixes $[\gamma_{ij}]$ defined as:

$$\begin{pmatrix} T_2 \\ Q_2 \end{pmatrix} = \begin{bmatrix} \gamma_{11} & \gamma_{12} \\ \gamma_{21} & \gamma_{22} \end{bmatrix} \begin{pmatrix} T_1 \\ Q_1 \end{pmatrix}. \qquad (2)$$

We assume that the positive directions of the heat fluxes are as shown by arrows in Figure 2a. The admittance and transmission matrices are related as:

$$[\gamma_{ij}] = \frac{1}{\alpha_{12}} \begin{bmatrix} -\alpha_{11} & 1 \\ -\det(\alpha) & \alpha_{22} \end{bmatrix}, \qquad (3)$$

where det ($\alpha$) denotes determinant of the matrix $[\alpha_{ij}]$. The inverse relation is obtained from Equation (3) replacing $\alpha$ with $\gamma$ and vice versa.

We start with writing the equation for the total admittance matrix for an SThM probe, which we denote $[t_{ij}]$. For generality, we assume that temperatures and heat fluxes are harmonically oscillating over time $t$ with an angular frequency $\omega$: $\widetilde{\delta T} = \delta T \cdot e^{i\omega t}$ and $\widetilde{Q} = Q \cdot e^{i\omega t}$ (including the case $\omega = 0$), where $i = \sqrt{-1}$. The tilde accent sign "~" underscores time-harmonic oscillations, $\delta T$ and $Q$ are complex amplitudes, and $\delta$ denotes the deviation from the time-averaged value. For simplicity of derivations but without the loss of generality, we assume as well that the temperatures of the base, $T_{\text{base}}$, and sample, $T_{\text{s}}$, are anchored to the temperature of the environment, RT. With a sample and the root of the probe anchored to the ambient temperature, we have:

$$\begin{pmatrix} Q \\ Q' \end{pmatrix} = \begin{bmatrix} t_{11} & t_{12} \\ -t_{12} & t_{22} \end{bmatrix} \begin{pmatrix} \delta T \\ 0 \end{pmatrix}, \tag{4}$$

which is illustrated by the schematic in Figure 2b. At the input port (port 1), $Q$ and $\delta T$ are amplitudes of the heat generated in the probe and temperature at the temperature-sensing element of the probe, respectively. The output port (port 2) is an environment, that is, a general distributed port, and its specific location can be chosen arbitrary in the environment. (See Supporting Information 1 for an example of an alternative choice of port 2.) The values of the matrix elements with index 2 and the flux $Q'$ are dependent on the specific choice of the output port. All matrix elements as well as $Q$, $Q'$, and $\delta T$ are complex numbers. Equation (4) takes into account also that the networks are reciprocal, and hence, $t_{21} = -t_{12}$, because the network consists of linear passive components, that is, resistors representing thermal resistances and capacitors representing heat capacitances.

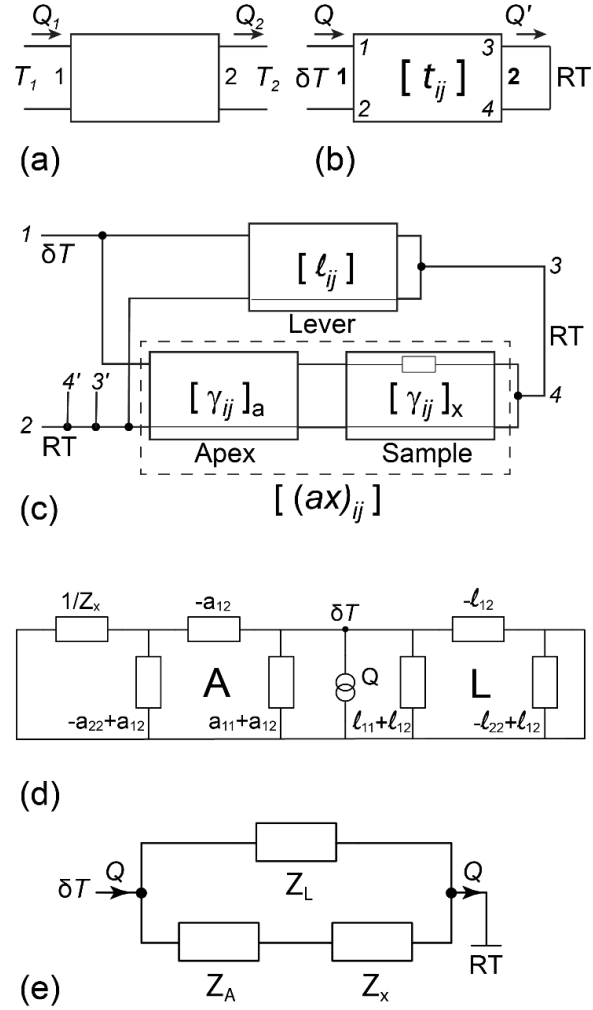

**Figure 2.** (a) Generic two-port four-terminals thermal network with heat fluxes $Q_{1,2}$ and temperatures $T_{1,2}$ at ports 1 and 2. (b) Generic two-port network describing an SThM probe in the active mode. The network is described by an admittance matrix $[t_{ij}]$. Ports are numbered in bold, and terminals are numbered in italics. RT stands for room temperature. (c) A thermal network for an SThM probe in active mode, where two-port networks for probe lever, probe apex, and sample are explicitly shown. The probe lever is described by an admittance matrix $[l_{ij}]$; probe apex with sample combined are described by an admittance matrix $[(ax)_{ij}]$, while the probe apex and sample are separately described by corresponding transmission matrixes $[\gamma_{ij}]_a$ and $[\gamma_{ij}]_x$; "x" stands for sample. See text for more details. 3' and 4' near the input terminal 2 denote an alternative port 2 (see Supporting Information 1). (d) π-equivalent representation of the circuit in (c). The values of admittances are expressed through respective admittance matrix elements. The sample is represented with one element, $1/Z_x$. "A" stands for apex, and "L" stands for lever. (e) Final simplified network for an SThM probe with generalized complex impedances as circuit elements. Impedances of the lever, apex, and sample are $Z_L$, $Z_A$, and $Z_x$, respectively. $Z_L$ and $Z_A$ are two of three calibration parameters of a probe. See text for details.

Next, we introduce admittance matrixes for parts of the probe: probe lever, $[l_{ij}]$, and the series chain of the probe apex and sample $[(ax)_{ij}]$; the probe-sample boundary is included into sample. The lever and the apex-sample chain are connected in parallel as depicted by the schematic in Figure 2c. For the parallel connection of the networks:

$$[t_{ij}] = [l_{ij}] + [(ax)_{ij}]. \tag{5}$$

Further, for transmission matrixes of the apex-sample chain and its elements (as denoted by subscripts "$a$" for apex and "$x$" for sample at corresponding transmission matrixes), it holds:

$$[\gamma_{ij}]_{(ax)} = [\gamma_{ij}]_x [\gamma_{ij}]_a, \tag{6}$$

from where it follows:

$$[\gamma_{ij}]_x = [\gamma_{ij}]_{(ax)} [\gamma_{ij}]_a^{-1}. \tag{7}$$

Since the probe-sample interaction volume is small inside the sample, the sample can be represented as a single lumped-element with a general complex impedance $Z_x$ connected in series as shown inside the sample box in the schematic in Figure 2c. The transmission matrix for a single series element with the impedance $Z_x$ is:

$$[\gamma_{ij}]_x = \begin{bmatrix} 1 & -Z_x \\ 0 & 1 \end{bmatrix}. \tag{8}$$

Equations (7) and (8) can be further used to construct the system of equations to derive probe calibration parameters and to express $Z_x$ through them. According to Equations (3) and (5):

$$[(ax)_{ij}] = [t_{ij}] - [l_{ij}] \xrightarrow{yields} [\gamma_{ij}]_{(ax)}, \tag{9}$$

and for apex:

$$[a_{ij}] \xrightarrow{yields} [\gamma_{ij}]_a, \tag{10}$$

where $[a_{ij}]$ is the admittance matrix for the apex. Since, $l_{21} = -l_{12}$, $a_{21} = -a_{12}$, matrices $[l_{ij}]$ and $[a_{ij}]$ contain in total six complex unknowns. Expressing component-wise $\gamma_{11,x} = 1$, $\gamma_{12,x} = -Z_x$, and $\gamma_{21,x} = 0$ through elements of $[l_{ij}]$, $[a_{ij}]$, and $[t_{ij}]$ yields a system of two equations to determine components of $[l_{ij}]$ and $[a_{ij}]$. The equation with $\gamma_{22,x} = 1$ is redundant since the sample's network with a single in-series component is obviously symmetric with $\gamma_{22} = \gamma_{11}$ (input and output are fully interchangeable).

In measurements, only the element $t_{11}$ can be evaluated. Therefore, the equations need to be expressed through $t_{11}$ and do not include other elements of the matrix $[t_{ij}]$. The elimination procedure leads to an equation:

$$Z_x = \frac{a_{11} + l_{11} - t_{11}}{a_{12}^2 + a_{22}(a_{11} + l_{11} - t_{11})}. \tag{11}$$

As seen, the sum $(a_{11} + l_{11})$ can be substituted with one unknown $y_{al}$: $y_{al} = a_{11} + l_{11}$, which eliminates one more unknown. To determine the three remaining unknowns—$a_{12}, a_{22}$, and $y_{al}$—one needs to perform three measurements with three known (calibration) samples, one of which can be represented by the probe out-of-contact with any sample.

Thus far, we did not include in the analysis the fact that temperature variations $\widetilde{\delta T}$ are not observed directly but determined indirectly by monitoring an output signal of a measurement circuit. Specifics of principle used for temperature measurement are not important at this point. However, a linear relationship between $\delta T$ and the output signal, that we denote $\delta v$, is assumed:

$$\delta v = \beta \cdot \delta T, \tag{12}$$

where $\beta$ is the proportionality coefficient. Equation (12) is valid as a small-signal, first-order approximation in any system. Furthermore, linearity of the response is pursued in the design of measurement systems. The impedance matrix $[t_{ij}]$ can be modified to be related to $\delta v$:

$$\begin{pmatrix} Q \\ Q' \end{pmatrix} = [t_{ij}] \begin{pmatrix} \delta v/\beta \\ 0 \end{pmatrix} = [t_{ij}/\beta] \begin{pmatrix} \delta v \\ 0 \end{pmatrix} = [t_{ij}^v] \begin{pmatrix} \delta v \\ 0 \end{pmatrix}, \tag{13}$$

where

$$t_{ij}^v = t_{ij}/\beta. \tag{14}$$

As seen, $\beta$ appears in the equations as the fourth unknown variable, which, however, cannot be independent from the other three variables because it is a factor at all matrix elements, and to determine $\beta$, an independent way to measure $\delta T$ is needed. However, in the calibration for the $Z_x$ measurement, another approach can be applied. Namely, a suitable equation for two of the variables $a_{12}$, $a_{22}$, and $y_{al}$ can be added to make $\beta$ one of independent variables. We set $a_{22} = a_{12}$, and substitute $a_{22}$ for $a_{12}$ in the equations. After replacement of elements of admittance matrices $[l_{ij}]$, $[a_{ij}]$, and $[t_{ij}]$ with their $v$-versions similar to Equation (13), we will obtain in place of Equation (11):

$$Z_x = \frac{1}{\beta} \frac{y_{al}^v - t_{11}^v}{a_{12}^v(a_{12}^v + y_{al}^v - t_{11}^v)}. \tag{15}$$

Noting the with the probe out-of-contact with a sample, $1/Z_x = 0$, we have the following set of three equations for the three unknowns $a_{12}$, $y_{al}$, and $\beta$:

$$\begin{cases} a_{12}^v + y_{al}^v - t_{11,\text{out}}^v = 0, \\ \beta Z_{c1} = \dfrac{y_{al}^v - t_{11,c1}^v}{a_{12}^v(a_{12}^v + y_{al}^v - t_{11,c1}^v)}, \\ \beta Z_{c2} = \dfrac{y_{al}^v - t_{11,c2}^v}{a_{12}^v(a_{12}^v + y_{al}^v - t_{11,c2}^v)}. \end{cases} \tag{16}$$

Here, $Z_{c1}$ and $Z_{c2}$ are known thermal impedances of two calibration samples, $t_{11,\text{out}}^v$, $t_{11,c1}^v$, and $t_{11,c2}^v$ are values of the matrix element $t_{11}^v$ with the probe out-of-contact with a sample, with the first and second calibration samples, respectively. To present the result for an arbitrary $Z_x$ after solving the system of equations, we introduce more convenient variables:

$$z_m = \frac{1}{t_{11,m}^v} = \frac{\delta v_m}{Q_m}, \tag{17}$$

where $m = 0$ corresponds to values measured with probe out-of-contact, and $m = 1$ and 2 correspond to values measured in contact with calibration samples 1 and 2, respectively; the values measured with an arbitrary sample are without a subscript. The final expression is:

$$Z_x = a\left(1 + \frac{b}{z_0 - z}\right) \tag{18}$$

with:

$$a = -\frac{(z_0 - z_2)Z_{c2} - (z_0 - z_1)Z_{c1}}{z_2 - z_1}, \qquad (19)$$

$$b = -\frac{(z_0 - z_2)(z_0 - z_1)(Z_{c2} - Z_{c1})}{(z_0 - z_2)Z_{c2} - (z_0 - z_1)Z_{c1}}. \qquad (20)$$

Besides, we obtain an expression for $\beta$:

$$\beta = \frac{(z_2 - z_1)\, z_0^2}{(z_0 - z_2)(z_0 - z_1)(Z_{c2} - Z_{c1})}. \qquad (21)$$

To better understand the meaning of the admittance matrix elements and the equivalent circuits after the variable elimination, we represent the two-port networks of the circuit in Figure 2c with their π-equivalents. The result is shown in Figure 2d with all resistive elements shown with their admittances expressed through values of the respective admittance matrix elements. We note that the value of the element $(-l_{22} + l_{12})$ is irrelevant because it is short-circuited. The value of the element $(-a_{22} + a_{12}) = 0$ because it was set that $a_{22} = a_{12}$. The elements $(a_{11} + a_{12})$, $(l_{11} + l_{12})$, and $-l_{12}$ are connected in parallel and can be replaced with one admittance, which we denote $1/Z_L$. Further, we set $a_{12} = -1/Z_A$, and arrive to a simple circuit shown in Figure 2e, where element values are shown with their thermal impedances: impedances of the lever, apex, and sample—$Z_L$, $Z_A$, and $Z_x$, respectively. With the probe out of contact, $1/Z_x = 0$, and, therefore, $z_0 = \beta Z_L$. With $Z_x = 0$, solving for the thermal impedance of the circuit in Figure 2e, $Z = z/\beta$, and with the use of Equation (18), we find:

$$b = \frac{-\beta\, Z_L^2}{Z_L + Z_A}. \qquad (22)$$

It is straightforward to check that from Equations (19)-(21):

$$a \cdot b \cdot \beta = z_0^2, \qquad (23)$$

and, hence:

$$ab = z_0 Z_L, \qquad (24)$$

and:

$$a = -(Z_L + Z_A). \qquad (25)$$

Using Equations (18), (21), (24), and (25), we also obtain:

$$Z_L = \frac{z_0}{\beta} = \frac{(z_0 - z_2)(z_0 - z_1)(Z_{c2} - Z_{c1})}{(z_2 - z_1) z_0}, \qquad (26)$$

$$Z_x + Z_A = \frac{z \cdot Z_L}{z_0 - z}, \qquad (27)$$

and:

$$Z_A = \frac{z_1 \cdot Z_L}{z_0 - z_1} - Z_{c1}. \qquad (28)$$

Equations (26)-(28) are the central result of this section. While the circuit in Figure 2e is identical to that usually employed for interpretation of the SThM probe signal, the values of *calibration parameters* $Z_L$, $Z_A$, and $\beta$ are assumed here to be obtained from the signals measured with calibration samples in configuration of a standard SThM imaging experiment and do not require knowledge about internal properties of a particular probe. All parameters in the equations are general complex numbers that assume a phase-sensitive detection of signals and that may not have straightforward relations with internal degrees of freedom of a probe.

In the next two sections, we verify calibration Equations (26)-(28) with the help of Finite-Elements Analysis (FEA) modeling of SThM probes of a commercially available type.

### 2.3. Linearized finite-elements model of an active SThM probe

The most employed SThM implementation to date is based on thermoresistive probes, where the temperature sensing and heating functions are combined in one element. To verify the calibration procedure presented in the previous section, we selected resistive probes KNT-SThM-2an (KNT) of Kelvin Nanotechnology (Glasgow, UK). The KNT probes are silicon nitride-based, cantilever-shaped, and compatible with AFM platforms. In the KNT probes, the heat is supplied as a Joule heat generated by an electrical current flow through a thin-film Pd strip near the probe apex, and the very same current is used to detect changes in the electrical resistance of the Pd strip resulting from its heating or cooling due to probe contact with a sample. The electrical resistance $R$ of the Pd sensing strip is considered to be a linear function of its temperature:

$$R = R_0(1 + \alpha \cdot \Delta T), \qquad (29)$$

where $R_0$ is the sensor resistance at a reference temperature, and $\Delta T$ is the sensor temperature change in respect to the reference temperature. The parameter $\alpha$ is the thermal coefficient of electrical resistance (TCR).

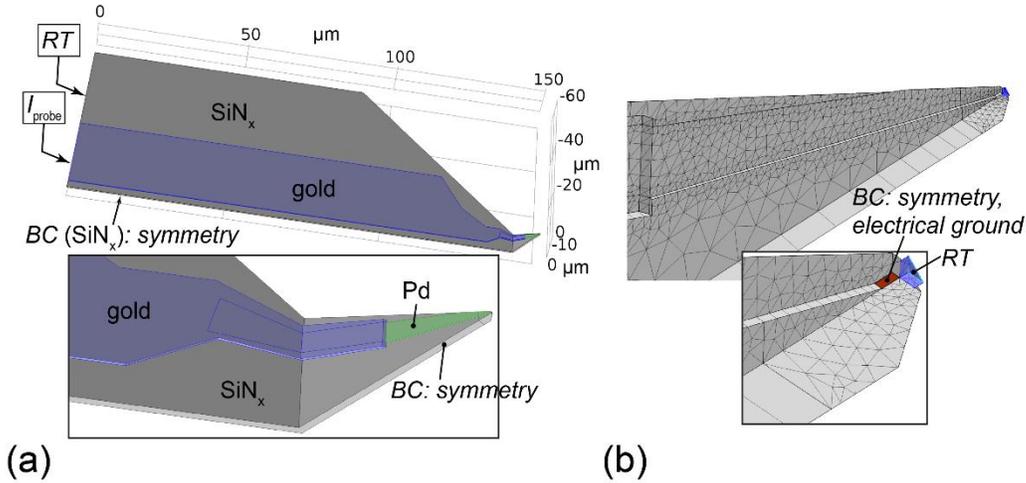

**Figure 3.** (a) Layout of the probe in the FEA model. $I_{probe}$ indicates the terminal face where the probe current is injected. The inset shows the magnified tip part of the probe. (b) Portion of the sensing tip of the probe in the FEA model that contains the Pd sensing strip, with mesh shown. The inset is the zoomed-in tip apex with the "thermal resistor" at the tip-sample contact face, which was used to model a sample. BC stands for boundary condition.

The layout of the FEA model of the KNT probe with materials and boundary conditions is displayed in Figure 3. Details of the model are described in the *Experimental section*. The model for the tip-sample contact is shown in the zoomed-in image of the tip apex in the inset of Figure 3b. To model the sum of a sample thermal spreading resistance and the probe-sample boundary resistance (= $R_{bs}^{th}$ in Figure 1b), a small domain—*"thermal resistor"*— interfaced with the contact area of the probe tip apex was introduced as shown in the inset in Figure 3b to model a sample. The boundary condition on the face of the domain opposite to the probe contact area was set to RT. The thermal conductivity of the material in the domain was varied as a model parameter to represent a certain, set, value of $R_{bs}^{th}$ by the thermal resistance of this domain between its outer face and the probe apex contact area.

In AC measurements without a DC current component, the generated heat is proportional to the square of the current amplitude, and information about the sensor resistance associated with temperature oscillations is contained in the voltage at the third harmonic of the frequency of the probe current.[43] Apparently, the probe is a non-linear system in respect to the input and output parameters—probe current amplitude and variations of the probe electrical resistance. To numerically model the AC probe response, time-dependent calculations are usually implemented,[64, 65] where the input current is applied as a sinusoidal function of time and, correspondingly, the probe response is calculated as a function of time. The resistance variations are extracted as the third harmonic of the calculated voltage drop across the probe. Such a computational approach is reliable with a sufficiently large number of points in a current oscillations period (>120 points), but it is time-consuming and is hardly suitable for studies involving sweeps of multiple parameters. Also, post-processing of the results of such calculations is not straightforward if, for example, distributions of amplitudes of temperature oscillations and heat fluxes in the probe are of interest. Therefore, in this work, we implemented problem linearization to take advantage of linear frequency-domain calculations with harmonic excitations, where explicit time-dependence is removed from equations for a single frequency via the standard separation of variables. After linearization, equations are solved numerically as stationary equations for complex amplitudes of temperature with frequency being a parameter.

To this aim, for AC calculations, we employed Joule heating interface in COMSOL Multiphysics with the preset *Thermal Perturbation, Frequency Domain* solver sequence setup. Due to the nonlinearity of the SThM probe-sample system, it is not possible to apply this solver sequence with an AC current as a perturbation because, in such a setup, thermal effects caused by Joule heating are not calculated since they are outside of the linearity domain and are not at the frequency of the perturbation (i.e., of the AC current). However, effects of a thermal perturbation on the current flow can be calculated since variations of resistance are linear in respect to temperature and, hence, heat oscillation amplitude.

To implement the linearization, for modeling the probe response with an AC current of a constant amplitude $I_0$ and angular frequency $\omega$, $I_{AC} = I_0 \cos(\omega t)$, calculations were performed in three steps. (I) First, a DC current $I'_{DC} = I_0/\sqrt{2}$ is applied in the probe model to "pre-heat" the model. The solution in this state serves as a linearization point for the other two steps. (II) As the second step, a time-harmonic, AC, current of an amplitude $I'_{AC} = I_0/\sqrt{2}$ and angular frequency $1\omega$ is added to the $I'_{DC}$ as a perturbation, and distribution of the Joule heat component oscillating at the frequency $\omega$ is calculated and extracted. This procedure yields the amplitude of the oscillating part of the Joule heat for $I_{AC} = I_0 \cos(\omega t)$ with $I_{DC} = 0$ (see Supporting Information 2 for more details and the rationale behind this choice of $I'_{DC}$ and $I'_{AC}$ values), and (III) at the last, third, step, the Joule heat amplitude distribution obtained at the previous step is applied as the time-harmonic perturbation with heat sources of the frequency $2\omega$. The third step yields the amplitude of the temperature oscillations and corresponding oscillations of the probe electrical resistance, both are at $2\omega$, when the current $I_{AC} = I_0 \cos(\omega t)$ is passed through the probe. Stepping through the calculation sequence is automated with the use of a script, and this approach required sufficiently less processor time and memory for calculation of the probe frequency response spectra compared to the time-dependent studies. Moreover, distributions of oscillation amplitudes and phases for different parameters over the probe model are calculated automatically and can be visualized directly with plots. (See Supporting Information 3 for examples of maps of simulated temperature distributions and Supporting Information 4 for examples of animated maps of temperature oscillations at different frequencies.)

Since the perturbation studies assume approximations in spatial distribution of current density, electrical resistivity, as well as in their temporal variations, we have verified the results obtained in this way by comparison with results of time-dependent studies and analyzing the transition to the static, DC, solution at small frequencies. The static solutions were obtained using a fully coupled *Electric Current-Heat Transfer* static study solver setup.

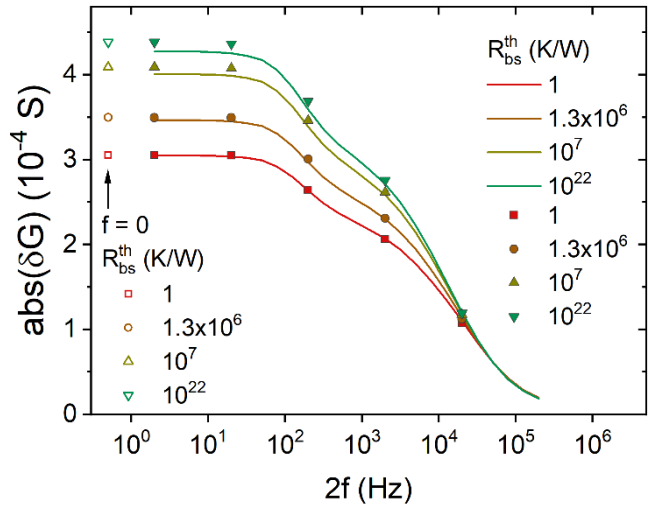

**Figure 4**. Frequency spectra of the probe conductance amplitude, $\mathrm{abs}(\delta G)$, calculated with the linearized model in the frequency domain (solid lines), as well as results of time-dependent (filled symbols) and static (empty symbols) solutions for several values of $R_{bs}^{th}$ as indicated in the plot legend. The variable $f$ in the plot argument is the probe current frequency.

For comparison of results of linearized, time-dependent, and static solutions, we used probe electrical conductance, rather than resistance, as a more accurate parameter in our modeling with a constant current amplitude. From Equation (29) for the temperature dependence of the probe electrical resistance $R$:

$$R = R_0(1 + \alpha \cdot Z \cdot RI^2), \tag{30}$$

where $I$ is the probe current and $Z$ is the total thermal impedance of the probe. As seen, $R$ enters both the left and right sides of the equation. Repeatedly substituting $R$ on the right side with its expression from Equation (30), we have:

$$\begin{aligned}
R &= R_0(1 + \alpha Z R_0(1 + \alpha Z R I^2)I^2) \\
&= R_0(1 + \alpha Z R_0 I^2 + (\alpha Z R_0 I^2)^2 + \cdots + (\alpha Z I^2)^N R_0^{N-1} R + \cdots) \\
&= R_0 \sum_{k=0}^{\infty} (\alpha Z R_0 I^2)^k = R_0 \frac{1}{1 - \alpha Z R_0 I^2},
\end{aligned} \quad (31)$$

or

$$G = \frac{1}{R} = G_0 - \alpha Z I^2. \quad (32)$$

where $G_0$ is the conductance at $I = 0$. For an AC current with a frequency $\omega$, these yields:

$$G = G_0 - \alpha Z_0 \frac{I_0^2}{2} - \alpha Z_{2\omega} \frac{I_0^2}{2} \cos(2\omega t), \quad (33)$$

where $Z_0$ and $Z_{2\omega}$ are thermal impedances at zero frequency and the frequency $2\omega$, respectively. As seen, the electrical conductance, $G$, has only zero-frequency and second-harmonic terms that are directly expressed through the AC current amplitude.

Figure 4 displays for comparison frequency spectra of the probe conductance amplitude, $\delta G$, calculated with the linearized model (solid lines), as well as results of time-dependent and static solutions for several values of $R_{bs}^{th}$. Overall, spectra shapes are similar to those reported earlier for the frequency response of the probe resistance.[68] The difference between the results obtained with linearized model and those obtained with direct calculations is small—about 2.5% or less. In the following, we use only results obtained with the linearized FEA models.

## 2.4 Verification results

To validate the proposed calibration procedure, we performed calculations of the probe response at different frequencies using the model with different set values of $R_{bs}^{th}$, and then applied Equations (26)-(28) to calculate the "measured" values of $R_{bs}^{th}$ from the calculated responses. The result of this verification is presented in Figure 5, where open symbols represent the function $y = x$ (identity between the function value and its argument). The frequency range of the AC current covered by the calculations spans between 1 Hz and 100 kHz. In the calculations, we used $R_{bs}^{th} = 5 \times 10^6$ K/W as $Z_{c1}$ and $R_{bs}^{th} = 10^7$ K/W as $Z_{c2}$. As obvious, the values calculated with the use of the calibration perfectly reproduce the set values of $R_{bs}^{th}$ for all frequencies. This verifies the calibration expressions, and this is in contrast to the results of Ref. [73]

where $R_{bs}^{th}$ values were calculated from model probe responses based on the currently adapted approach, even though the probe equivalent circuit used was topologically identical to that in Figure 2e.

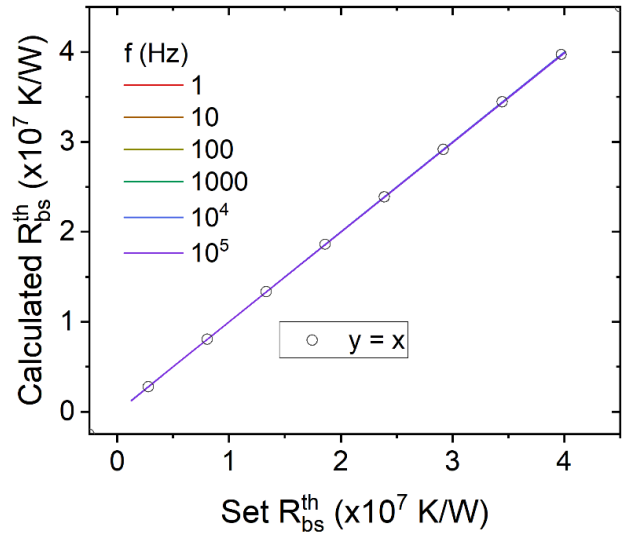

**Figure 5.** $R_{bs}^{th}$ calculated from the probe responses simulated in the FEA models vs. $R_{bs}^{th}$ set in the models. Calculations were performed following the proposed calibration procedure as described in the text. Solid lines are calculations for different probe current frequencies indicated in the plot legend; plot lines for different frequencies coincide. Empty circles correspond to the identity function $y = x$.

**2.5 Ways to practical implementation**

The proposed calibration approach requires suitable calibration samples. Any calibration measurements have to satisfy several requirements. They should be accurate, reproducible, and traceable, and they should yield identical results in different laboratories. Apparently, for that, calibration samples should be identical, hence, accurately fabricated and characterized. Furthermore, they have to be convenient to use. Therefore, creation of such samples (the sample design, fabrication and characterization) is a standalone, independent, task, which remains outside the scope of our work. In this section, we discuss how it can be realized in practice.

Overall, three parameters need to be determined for measurements of an arbitrary sample: $Z_L$, $Z_A$, and probe-sample contact radius $r$. With knowledge of $Z_L$, the parameter $\beta$ can be determined with the use of the out-of-contact signal $z_0$ and Equation (26). However, knowledge of $\beta$ is generally not needed to determine sample properties. The proposed approach assumes calibration measurements with samples offering different and known thermal impedances $Z$ (at each frequency), and in principle, three samples can be used to calibrate the three unknowns for a given probe.

In the consideration above, we assumed that $Z_x = R_s^{th} + R_b^{th}$ (see Figure 1b). Hence, both $Z_{c1,2}$ in Equations (26)-(28) include the probe-sample thermal boundary impedance, which is in series with $Z_A$ and unknown, depending on sample material and conditions in the probe-sample contact. This prevents straightforward determination of $Z_A$ with the use of Equation (28). Still, we note that in Equation (26), $Z_L$ is dependent on the difference $(Z_{c2} - Z_{c1})$, but not on the absolute values of $Z_{c1,2}$. Therefore, $Z_L$ and $r$ can be calibrated without knowing the boundary resistance $R_b^{th}$, provided that $R_b^{th}$ is the same for calibration samples.

Calibration samples must satisfy also some additional conditions, since $Z_x$ is a function of $r$, and $r$ is one of the parameters to be calibrated. This can be seen, for example, with bulk samples, for which the spreading impedance:

$$R_s^{th} = 1/4k_s r \qquad (34)$$

where $k_s$ is the sample material thermal conductivity. Let us suppose that two different samples have different $k_s$. Substituting $Z_{c1,2}$ in Equation (26) for $R_s^{th}$ with different values of $k_s$, we find that $Z_L \sim 1/r$, and $Z_L$ and $r$ cannot be separated (i.e., determined independently). Therefore, with measurements with bulk samples, the contact radius needs to be calibrated in a different, dedicated, setup. (Examples of measurements and calibration of $r$ in dedicated setups can be found in literature [74, 75].) Hence, ideally, for calibration of $Z_L$ and $r$ in one measurement setup, the relation between $R_s^{th}$ and $r$ for calibration samples should be such that calibration measurements yield an equation system, which can be solved in respect to $Z_L$ and $r$, separating them.

Fabrication and application of samples satisfying the conditions of the two previous paragraphs were reported in recent publications by Spièce et al.[36] and by Gonzalez-Munoz et al.[31], where the authors employed as a reference sample a structure made of a $SiO_2$ layer formed on a surface of a bulk Si substrate. The expressions for the spreading resistance of a circular contact on top of a layer that coats an infinite half-plane substrate were derived in by Dryden[76] and such that of $Z_L$ and $r$ can be separated measuring samples with different layer thicknesses. $SiO_2$/Si structures are well suited for this purpose because thermal properties of both $SiO_2$ and Si are well known, and the layer thickness can be accurately measured. However, the calculation of the structure spreading resistance involves integration or summation of infinite series (see Refs. [31, 76-78] for expressions), therefore, calculations to determine probe calibration parameters can be performed only numerically. Besides, the thermal resistance of the interface between the oxide layer and Si substrate is unknown and generally needs a separate measurement. Addressing these complications, Gonzalez-Munoz et al.[31] fabricated the $SiO_2$/Si structure as a wedge with linearly varying layer thickness and measured the SThM probe signal as a function of layer thickness. Then, using an analogue of Equation (27), they fit the analytically derived probe response to the experimentally measured one to obtain the probe-sample contact radius, SiO$_2$/Si interface thermal resistance, and a factor $c_e$, that the authors called probe "sensitivity" or "tooling factor". The "tooling factor" appeared in the expression (Equation (1) in Ref. [31], which we reproduce below as Equation (35)) that the authors used to determine the probe-sample thermal resistance $R_x$ as a function of local thickness, $t$, of the SiO$_2$ wedge-like layer:

$$R_x(t) = c_e \frac{V_x(t) R_p}{V_{nc} - V_x(t)} \tag{35}$$

Equation (35) is nearly identical to Equation (27) with correspondences $(Z_x + Z_A) \to R_x$, $z \to V_x$, $z_0 \to V_{nc}$, $Z_L \to (c_e \cdot R_p)$ and the same physical meaning of the corresponding parameters. The factor $c_e$ was introduced by the authors to account for the difference between the measured $R_x(t)$ values and the values obtained straightforwardly with the probe thermal resistance, $R_p$ (an analogue of $R_l^{th}$ in the circuit of Figure 1b), calibrated with the currently adapted approach when the distributed nature of the probe is neglected. As is clear, the separate calibration of $R_p$ is an unnecessary step because the full product $(c_e \cdot R_p)$ can be determined and used as one, single, parameter. With their method, the authors also determined the contact radius of the probe and value of the sum $(R_b^{th} + Z_A)$. As can be concluded, the design of the measurements by Gonzalez-Munoz et al. in Ref. [31] is such that it experimentally realizes the probe calibration approach proposed here, however, in a somewhat different form. While the measurements were made at DC, it is clear that the SiO$_2$/Si calibration samples can be straightforwardly used at AC as well. It is also clear that that layer-on-bulk calibration samples can be made in different configurations, for example, as a series of well-defined SiO$_2$ steps on a flat Si substrate like those used by Guen et al. in Ref. [79].

Next, we turn to the probe parameter $Z_A$. Experimental determination of $Z_A$ poses significant difficulties and remains a largely unsolved problem of SThM. Determination of $Z_A$, which is an analogue of $R_t^{th}$ in Figure 1b, largely relies on modeling, which requires knowledge of probe internal properties. This is because in measurements, $Z_A$ is a part of the sum $(R_b^{th} + Z_A)$, where the boundary resistance $R_b^{th}$ is a function of poorly controlled sample surface conditions, such as surface roughness and presence of contaminations. It is generally dependent on the sample material and unknown. Measurements with calibration samples as described in the previous paragraph generally return the sum $(R_b^{th} + Z_A)$ in place of $Z_A$ in Equation (28) with a $R_b^{th}$ value strictly valid for the probe-calibration sample pair only. However, useful information about sample material can be obtained under conditions that $R_b^{th}$ remains sufficiently constant during measurements with other parameters varied in a known way.[31, 36, 79-83] In the next section, we return to this issue in connection with measurement sensitivity.

**2.6 Analysis of probe sensitivity**

In this section, as an example of application, we use the calibration parameters to analyze probe sensitivity and consider whether the sensitivity of a SThM probes to variations of sample thermal conductivity can be increased by tuning probe parameters. The analysis is based on the equivalent circuit in Figure 2e with parameter values assumed to be determined through calibration with Equations (26)-(28).

We start our analysis noting that the probe temperature increase (excess temperature) $\Delta T$ cannot be arbitrary large, even though the temperature increase straightforwardly increases the output signal. The value of $\Delta T$ must be limited by a maximum value, which is safe for the probe and sample.

*Resistive probes with combined heating and sensing functions*

For certainty, we continue the analysis with the KNT resistive probes. In the case of resistive probes, with increasing electrical resistance of the sensing strip at a fixed probe current, the temperature of the sensing strip will increase, and the temperature variations of the strip due to the contact with the sample will increase accordingly. Since the excess temperature of the probe should be limited, the maximal probe current should be decreased to keep the temperature under the limit, which reduces the output voltage that is proportional to the current. The same is true with an increase of the probe lever thermal resistance, $R_l^{th}$ (Figure 1b), which is responsible for the probe temperature when the probe is out of contact with a sample. Note that $R_l^{th}$ is an analogue of $Z_L$ (Figure 2e). Without loss of generality, we limit our further consideration to constant current with conclusions directly valid at low frequencies AC. At higher frequencies, probe temperature oscillations decrease, and excess temperature is determined by the average power deposited into a probe, limiting the AC amplitude similarly to DC value.

To compare sensitivity of two arbitrary probes, we set a probe-independent limiting excess temperature at the sensing element of the probe to a value that we denote $\Delta T_{max}$, assuming that this temperature is achieved at the probe operational current $I$:

$$\Delta T_{max} = I^2 R_0 Z_L = const. \tag{36}$$

From Equation (30) with $R \approx R_0$, the corresponding probe response, expressed as a variation of the voltage drop across the probe due to the change in the resistance, is:

$$\delta V = \delta R \cdot I = \alpha Z I^2 R_0^2 \cdot I = \alpha R_0^{1/2} \, \Delta T_{\max}^{3/2} \frac{Z}{Z_L^{3/2}}. \tag{37}$$

We define the *absolute* probe sensitivity $S$ as the absolute derivative of the probe response $\delta V$ over the sample thermal impedance $Z_x$ under the condition, when $\Delta T_{\max}$ is reached:

$$S = \frac{d(\delta V)}{dZ_x} = \alpha R_0^{1/2} \, \Delta T_{\max}^{3/2} \cdot \frac{d\left(Z/Z_L^{3/2}\right)}{dZ_x}. \tag{38}$$

Interpretation of the factor $\alpha R_0^{1/2}$ in the last equation is straightforward with some reservations associated with for the distributed nature of the probe. The parameter $\alpha$ should be understood as effective and generally different in value from the TCR of the material of the sensing element of the resistive probe. The value of $R_0$ can be also different from that determined by a direct measurement. This means that alterations of the layout of the sensing strip of the probe can influence both $R_0$ as well as $\alpha$.

The last factor in Equation (38) can be calculated using the thermal equivalent circuit in Figure 2e. With:

$$Z = \frac{Z_L(Z_A + Z_x)}{Z_L + Z_A + Z_x} = Z_L - \frac{Z_L^2}{Z_L + Z_A + Z_x}, \tag{39}$$

the derivative is:

$$\frac{dZ}{Z_L^{3/2} dZ_x} = \frac{Z_L^{1/2}}{(Z_L + Z_A + Z_x)^2}. \tag{40}$$

In this formulation, to get the correct result in respect to sample thermal conductivity, $Z_x$ should not include the probe-sample boundary resistance. Hence, we redefine $Z_A$ by including the boundary resistance in this term. We will denote the redefined $Z_A$ as $Z_{Ab}$:

$$Z_{Ab} = Z_A + R_b^{th}, \tag{41}$$

and $R_s^{th}$ should be further substituted for $Z_x$ in expressions.

Analyzing the probe sensitivity factors for the optimization purpose further, we have to take into account the method used for measurement of the probe response as well as the sample properties. In the case, when the probe response is measured with the use of a Wheatstone bridge, that is, with a differential method, the sensitivity should be defined through the absolute change of the circuit signal, rather than related (normalized) to a response at a certain state, such as the out-of-contact state. The signal normalization would eliminate the product $\alpha R_0^{1/2} \Delta T_{\max}^{3/2}$ from the equation, while, obviously, all the factors in the product directly affect the sensitivity. On the other side, when imaging the sample thermal conductivity, $k_s$, it is expected that the variation of $k_s$ are approximately proportional to the average thermal conductivity in the imaged area. In such a case, the change of the sample thermal conductivity in the derivative in Equation (40) should be related to the thermal conductivity itself. With $R_s^{th} = 1/4k_s r$, $dk_s/k_s = -dR_s^{th}/R_s^{th}$, and we may continue operating with $R_s^{th}$, without switching to $k_s$, in the equations below to simplify notation.

The sensitivity $S_{Rs}$ defined in respect to the *relative change* of the sample thermal conductivity under the condition, when $\Delta T_{\max}$ is reached, then:

$$S_{Rs} = \frac{R_s^{th} \cdot d(\delta V)}{dR_s^{th}} = \alpha R_0^{1/2} \Delta T_{\max}^{3/2} \cdot \frac{R_s^{th} Z_L^{1/2}}{(Z_L + Z_{Ab} + R_s^{th})^2}. \tag{42}$$

Note that the relative sensitivity is a dimensional quantity showing the change of $Z$ expressed in K/W per a fractional change in $R_s^{th}$. Figure 6 displays a 3D plot of the function $f_Z$, which is the last factor in Equation (42) and depends only on thermal impedances:

$$f_Z = \frac{R_s^{th} Z_L^{1/2}}{(Z_L + Z_{Ab} + R_s^{th})^2} \tag{43}$$

with $Z_L$ and $R_s^{th}$ as independent variables and $Z_{Ab} = 10^7$ K/W.

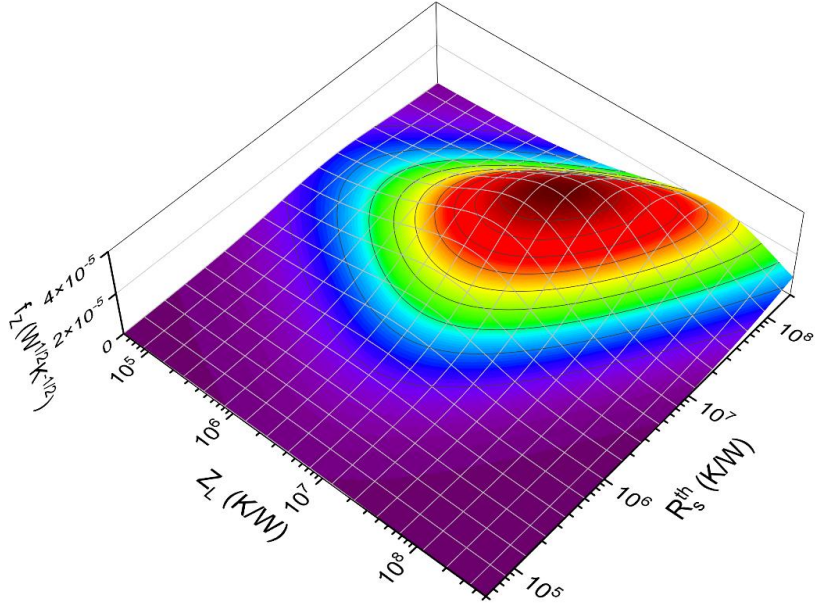

**Figure 6.** 3D plot of the function $f_Z$ defined by Equation (43) vs. $Z_L$ and $R_s^{th}$ with $Z_{Ab} = 10^7$ K/W.

The function has a global maximum at $(Z_L = Z_{Ab}, R_s^{th} = 2Z_{Ab})$ with a value $f_{Z,max} = Z_{Ab}^{-1/2}/8$. Since in the SThM practice mainly $R_s^{th} < Z_{Ab}$, the maximum is not reached. Still, for a given, fixed $R_s^{th}$ and $Z_{Ab}$, the function has a maximum in respect to $Z_L$ at:

$$Z_L = (Z_{Ab} + R_s^{th})/3, \qquad (44)$$

which points to a possibility of optimization of the probe sensitivity by tuning its $Z_L$ value for measurements in a certain range of the sum $(Z_{Ab} + R_s^{th})$. The value of the function $f_Z$ at a maximum along a constant $R_s^{th}$ is:

$$f_{Z,max}(R_s^{th}, Z_{Ab} = const) = \frac{3\sqrt{3}\, R_s^{th}}{16(Z_{Ab} + R_s^{th})^{3/2}}. \qquad (45)$$

To estimate how close to optimal the KNT probe parameters are, we inspect the values for the probe thermal resistance for the KNT probes reported in recent literature. As discussed in the previous section, in a form closest to the definition of $Z_L$ here, this parameter is given by Gonzalez-Munoz et al. in Ref. [31] for DC, where a probe resistance of $R_p = 1.61 \times 10^5$ K/W is in a product with a "tooling factor" $c_e \approx 8.29$, which yields $Z_L = c_e R_p = 1.33 \times 10^6$ K/W. In turn, in Ref. [73], while $Z_L$ as defined here was not used, the probe resistance values were obtained with the use of a calibrated FEA model and fall in the range from 6.3×10⁵ K/W to 8.8×10⁵ K/W. In both the reports, the sum $(Z_{Ab} + R_s^{th})$ is close to a few 10⁷ K/W. Hence, with the $Z_L$ value close to 10⁶ K/W, the probe $Z_L$ was somewhat suboptimal for the measurements.

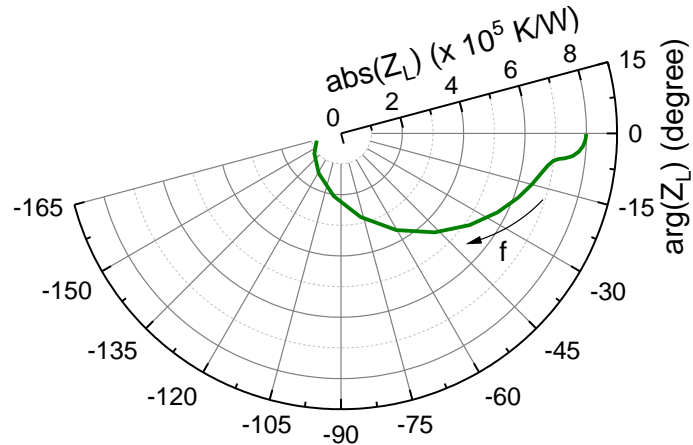

**Figure 7.** Nyquist plot of the probe calibration parameter $Z_L$ calculated with the FEA model for the standard KNT probe in a range of the probe current frequency, *f*, from 1 Hz to 100 kHz.

To find a way to optimize the probe, we employed Nyquist plot representation for the complex, frequency-dependent values of the probe parameters $Z_L$, $Z_A$, and $\beta$. The Nyquist plot of $Z_L$ for the standard probe simulated with the FEA model (Section 2.3) is shown in Figure 7 (in polar coordinate system). The plot reveals two semicircular arcs with significantly different radii. The shape of the plot suggests that the probe $Z_L$ can be represented with two R||C circuits connected in series ("R||C" denotes a resistor and a capacitor connected in parallel). The radius of an arc in the plot is determined by the resistance in the R||C circuit. Therefore, the lower frequency semicircle, that has a smaller radius, corresponds to a small thermal resistance in parallel with a relatively large thermal capacitance, while the higher frequency, large, semicircle corresponds to a larger thermal resistance in parallel with a smaller thermal capacitance. Looking at the probe shape (Figure 3a), it becomes clear that the small semicircle corresponds to the (massive) probe lever, while the large one corresponds to the (small) probe sensing tip. This is in agreement with the lumped-element KNT probe model constructed by Bodzenta et al. in Ref. [68]. Hence, $Z_L$ is determined mainly by the thermal resistance of the probe tip.

Since the sensing strip is located at the very apex of the tip, where the tip is narrowest and, hence, has larger thermal resistance, it can be assumed that a slight modification of the near-apex part of the probe may lead to a noticeable change of the probe performance with alteration of the value of $Z_L$. To check this assumption, we have modified the shape of the sensing strip of the probe in our FEA model. The modification is illustrated in Figure 8a. Namely, the crossing point of the internal edges of the Pd sensing strip was moved by about 1.7 μm closer to the probe apex. As is clear, with this modification, the heat generation is moved towards the tip apex.

The simulated Nyquist plots for $Z_L$ and $Z_A$ for both the standard and modified probes in the frequency range from 1 Hz to 100 kHz are show in Figure 8b. As seen, as a result of the simple modification, $Z_L$ is increased and $Z_A$ is reduced, apparently, due to shift of the strip electrical resistance and the center of gravity of the hot spot on the probe tip towards the tip-sample contact. This result indicates that properties of the very apex of the tip overwhelmingly contribute to the $Z_L$ and $Z_A$ of the probe.

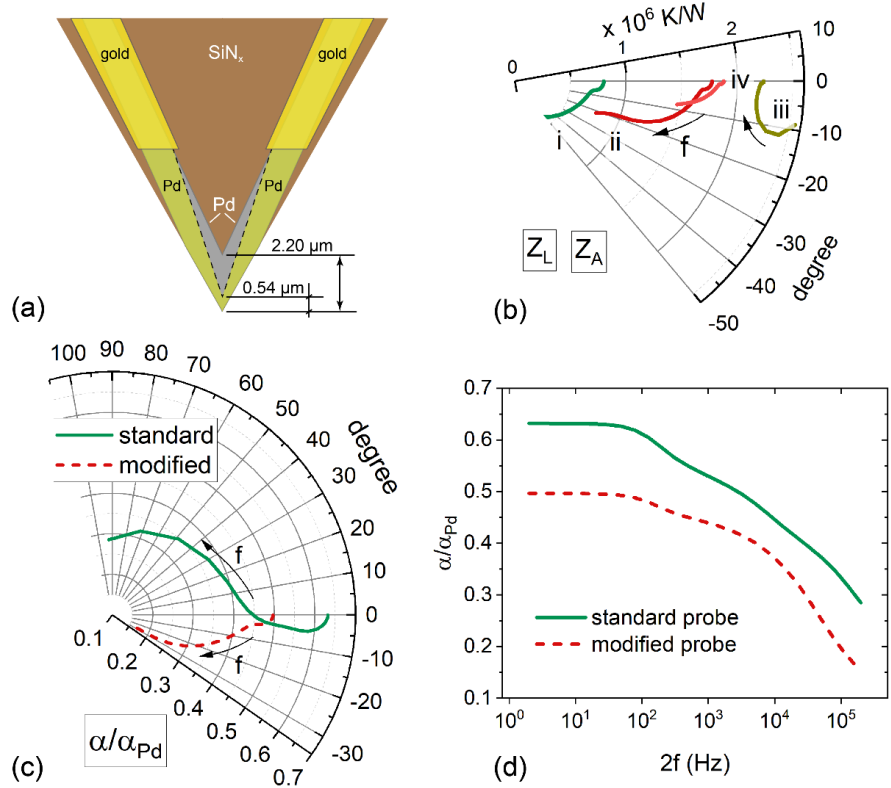

**Figure 8.** (a) A sketch showing the modification of the probe sensing strip. The removed part of the original strip as in the standard probe is seen in gray, above the dashed line. (b) Calculated Nyquist plots of the parameters $Z_L$ and $Z_A$ of the standard and modified probes in the frequency range from 1 Hz to 100 kHz: (i) $Z_L$ of the standard probe, (ii) $Z_L$ of the modified probe, (iii) $Z_A$ of the standard probe, (iv) $Z_A$ of the modified probe. (c) Nyquist plots of the calculated ratio $\alpha/\alpha_{Pd}$ for standard (solid line) and modified (dashed line) probes in the frequency range from 1 Hz to 100 kHz. Parameter $\alpha$ is calculated using the calibration parameter $\beta$ and Equation (48). (d) Amplitude Bode plots of the calculated ratio $\alpha/\alpha_{Pd}$ for standard (solid line) and modified (dashed line) probes represented versus doubled probe current frequency. The variable $f$ in plots is the probe current frequency.

The FEA models for the two probes showed a larger relative sensitivity of the modified probe in the $R_{bs}^{th}$ range near $10^7$ K/W, as was determined directly from the calculated probe response as a function of $R_{bs}^{th}$. The larger sensitivity is expected because of a larger $Z_L$ and smaller $Z_A$ of the modified probe. For the calculations, we applied the first equality in Equation (42) substituting $R_s^{th}$ with $(R_{bs}^{th} - R_b^{th})$, where $R_b^{th} = const$, to have $R_{bs}^{th}$ as an independent variable as it is in the FEA models. The formula for the sensitivity in the FEA model becomes, then:

$$S_{RS}^{FEA} = (R_{bs}^{th} - R_b^{th})\frac{d(\delta V)}{dR_{bs}^{th}}, \tag{46}$$

and $S_{RS}^{FEA}$ can be calculated numerically from a simulated $\delta V$ vs. $R_{bs}^{th}$ dependence.

Figure 9a displays plots of $S_{RS}^{FEA}$ vs. $R_s^{th} = R_{bs}^{th} - R_b^{th}$ with $R_b^{th} = 10^7$ K/W for standard and modified probes as well as the ratio of the corresponding $S_{RS}^{FEA}$ values at the AC frequency of 1 Hz. As seen, for $R_s^{th}$ in the range between $10^4$ K/W and $10^6$ K/W the increase is by a factor $\approx 1.65$. Specifically, at $R_s^{th} = 10^5$ K/W, the factor is 1.65. This value is in a perfect match with the sensitivity determined from the expression of the probe response, $\delta V$, through the calibration parameters $Z_L$ and $\beta$:

$$\delta V = \beta \frac{Z}{Z_L} \Delta T_{max} \tag{47}$$

which can be obtained from Equations (36) and (37). At $R_s^{th} = 10^5$ K/W, the use of this expression yields the sensitivity increase factor of 1.66. This directly shows how calibration parameters can be used to compare the performance of probes. Note that in Equation (47), the exact principle of operation of the probe is hidden.

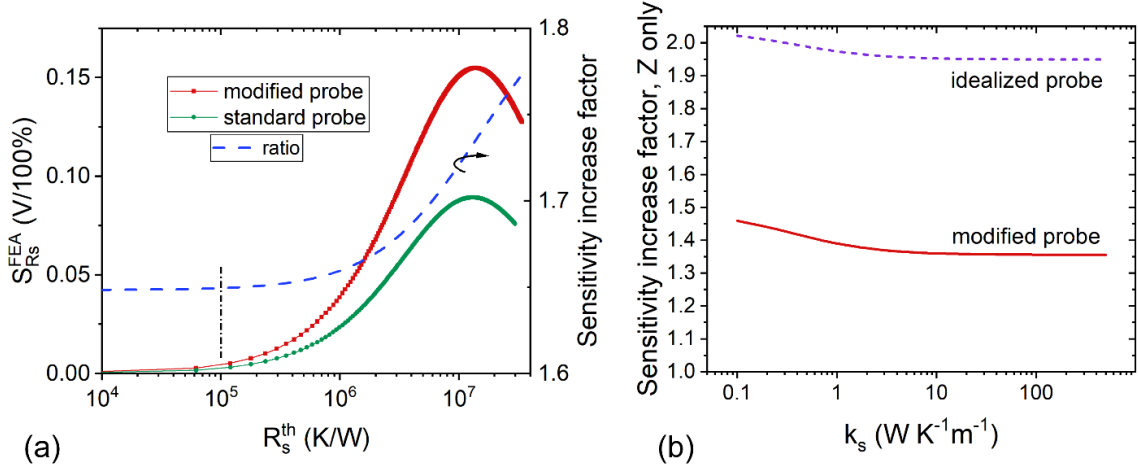

**Figure 9**. (a) Solid lines with symbols (left axis): plots of the probe relative sensitivity, $S_{RS}^{FEA}$, vs. $R_s^{th}$ calculated from the probe response in the FEA models with $R_b^{th} = 10^7$ K/W for standard and modified probes (red and green lines, respectively). Dashed line (right axis): sensitivity increase factor (i. e., $S_{RS}^{FEA}$ of a probe related to the $S_{RS}^{FEA}$ of the standard probe) for the modified probe at a probe current frequency of 1 Hz. (b) Sensitivity increase factors (in respect to sensitivity of the standard probe) as functions of sample thermal conductivity, $k_s$, for the modified (solid line) and idealized (dashed line) probes. The probe-sample contact radius is assumed to be 50 nm.

Analysis with Equation (42) shows that increase in the probe relative sensitivity due to the changes of $Z_L$ and $Z_A$ values alone is by a factor of 1.33. The rest, that is, an additional factor of about 1.2, is due to the increase of $R_0$ combined with a decrease in $\alpha$ (which plays the role of the TCR of the sensing strip material). To be specific, $R_0$ values at the linearization point in the FEA model are 113 Ω for the standard probe and 277 Ω for the modified probe. Parameter $\alpha$ can be calculated from $\beta$ determined via calibration. From the definition of the parameter $\beta$ in Equation (12), when it is applied to the resistive probes in the FEA models, and Equation (37), we deduce:

$$\beta = \alpha I R_0. \qquad (48)$$

In the FEA models, values of $\alpha$ determined with the use of this expression and calibrated $\beta$ are about 63% of $\alpha_{Pd}$ for the standard probe and about 49% of $\alpha_{Pd}$ for the modified probe at low frequencies with $R_0$ values taken at the linearization point.

It is of interest to examine the frequency behavior of $\alpha$. Values of $\alpha$ were calculated in the FEA models in the AC frequency range from 1 Hz o 100 kHz using calibrated $\beta$ and Equation (48) with the product $IR_0$ calculated as the voltage drop through the probe at the linearization point. The obtained (complex) values of the ratio $\alpha/\alpha_{Pd}$ are presented as Nyquist and amplitude Bode plots in Figure 8c and d, respectively. As seen, the absolute values of $\alpha$ are significantly below the TCR of the sensing strip material and decrease with increasing frequency. The shape of the dependence of $\alpha$ on frequency is closely resembling the frequency response spectra of the probes, which can be interpreted so that the thermal field along the sensing strip is determined by the heat propagation in its $SiN_x$ substrate.

Using the calibration parameters and Equation (43), we have also calculated the highest factor of relative sensitivity increase (ratio of sensitivities) that is possible by tuning only $Z$-parameters of the probe with other parameters fixed. We compared $f_Z$-function values (Equation (43)) for the standard probe and for an idealized probe with $Z_A = 0$ and $Z_L = (Z_{Ab} + R_s^{th})/3$ (Equation (44)) at low probe current frequencies. Figure 9b displays a plot of the sensitivity increase factor vs. thermal conductivity of the sample material, $k$, for a probe-sample contact radius $r = 50$ nm and a probe-sample boundary resistance of $10^7$ K/W ($\rho_i^{th} = 7.85 \times 10^{-8}$ K m² W⁻¹). As seen, the expected sensitivity of an idealized, impedance-matched probe is only about two times larger than the sensitivity of the standard KNT probe if only $Z$-parameters are considered. Figure 9b also shows for comparison the sensitivity increase factor for the modified vs. standard probe.

It is worth noting as well that the thermal noise generated by the probe will increase with increasing $R_0$ and probe temperature, $T$, as $R_0^{1/2} T^{1/2}$. Therefore, as can be deduced from Equation (42), at the Johnson–Nyquist (thermal) noise limit, the signal-to-noise ratio of the resistive probe is independent of $R_0$ and sublinear (weaker, than linear) in respect to $\Delta T_{max}$.

*Probes with separated heating and sensing functions*

The combined heating and sensing functions of the sensing element of the resistive SThM probe with the necessity to keep the sensing element temperature under a certain value dictate that the probe current "reading" the sensor element resistance must be decreased with increased $Z_L$ and $R_0$, which decreases the probe response (output voltage) value. This is reflected, in particular, by the presence of the factor $Z_L^{-3/2}$ in the right part of Equation (37), which is transferred into $Z_L^{1/2}$ in the expression for sensitivity, Equation (42). In this connection, it is of interest to note that for a probe with separated heating and temperature sensing functions, $Z_L$ will enter the nominator of the sensitivity function $f_Z$, Equation (43), as $Z_L$. In this case, the function $f_Z$ does not have a global maximum, but a maximum in respect to $Z_L$ at a fixed $Z_x$ still exists. Namely, starting with Equation (47), we get:

$$S_{Rs} = \frac{R_s^{th} \cdot d(\delta V)}{dR_s^{th}} = \beta \, \Delta T_{max} \cdot \frac{R_s^{th} \, Z_L}{(Z_L + Z_{Ab} + R_s^{th})^2}, \quad (49)$$

and

$$f_{Z,max}(R_s^{th}, Z_{Ab} = const) = \frac{R_s^{th}}{4(Z_{Ab} + R_s^{th})} \quad (50)$$

at

$$Z_L = Z_{Ab} + R_s^{th} \quad (51)$$

Assuming application of probes to measurement with identical values of $R_s^{th}$ and $Z_{Ab}$ and with the use of Equations (36), (42), (44), (45), and (48)-(51), we obtain:

$$\frac{S_{Rs}^{sep}}{S_{Rs}^{res}} = \frac{4}{3} \frac{\beta^{sep}}{\beta^{res}} \frac{\Delta T_{max}^{sep}}{\Delta T_{max}^{res}} \quad (52)$$

for probes with correspondingly optimized $Z_L$ values, where subscripts "res" and "sep" refer respectively to resistive probes and probes with separated heating and temperature detection functions.

*Boundary resistance and sensitivity*

As can be seen from plots in Figure 9 and Equation (52), the probe sensitivity in a specific task can be increased via optimization of the probe design, but the expected increase—few times—is not significant. Therefore, the main factor suppressing the sensitivity is relatively large values of $R_b^{th}$, which enters the denominator of the expression for sensitivity, Equation (42), through $Z_{Ab} = Z_A + R_b^{th}$ in the power of 2. Since $Z_A$ is only by a factor on the order of unity larger that the spreading resistance for the bulk material that the probe is made of, $Z_{Ab}$ is dominated by the probe-sample boundary thermal resistance $R_b^{th}$ for probes made of silicon nitride or silicon, which is usually the case. The boundary resistance cannot be made arbitrary small. Its value is fundamentally determined from below by the properties of the heat transfer across the interface in the probe-sample material pair, i.e., Kapitza resistance for the material pair [84].

For reliable measurements and imaging, it is necessary that $R_b^{th}$ is constant or its variations are smaller enough than the expected variations of the material spreading resistance $R_s^{th}$. If this condition is fulfilled, even if $R_b^{th}$ is large, meaningful measurements of sample thermal conductivity can be possible. In practice, such a condition can be realized with specially prepared nearly-uniform-material samples with clean and smooth surfaces, like semiconductor wafers.

A sample surface can be characterized by the probe-sample interfacial thermal resistance, $\rho_i^{th}$ (defined per unit area of the interface). For non-metals, the lower boundary for $\rho_i^{th}$ is about $10^{-8}$ K m² W⁻¹. The upper boundary for an interface between very dissimilar materials with clean surfaces is about $10^{-7}$ K m² W⁻¹.[85] However, in practice, it can be much larger. Since the lower boundary is about 10% of the upper boundary for clean surfaces, we may expect that variations of the $\rho_i^{th}$ along a sample surface without a special surface preparation can be a significant fraction of $10^{-7}$ K m² W⁻¹. Therefore, for estimates, we further assume that meaningful measurements are possible when the sample spreading resistance, $R_s^{th}$, and the boundary resistance, $R_b^{th}$, and, hence, their absolute variations across a sample, are approximately equal. $R_b^{th}$ is calculated from $\rho_i^{th}$ and probe-sample contact radius, $r$, as:

$$R_b^{th} = \rho_i^{th}/\pi r^2 \quad (53)$$

This dependence on $r$ is stronger that for the sample spreading resistance, which is $\sim 1/r$, Equation (34), and there is a contact radius value, at which $R_b^{th} = R_s^{th}$ for a given probe-sample pair. Taking $R_b^{th} = R_s^{th}$, we find the following relation:

$$\frac{k_s \rho_i^{th}}{r} \approx 1. \quad (54)$$

This expression can be used, for example, to estimate range of the material $k$-values when meaningful measurements can be anticipated for a probe with a given $r$ and an expected range of $\rho_i^{th}$. For instance, with $r = 100$ nm and $\rho_i^{th} \approx 10^{-7}$ K m² W⁻¹, it should be that $k_s < 1$ W K⁻¹ m⁻¹, which limits the application of the high-resolution SThM to samples with small thermal conductivities. With low enough sample thermal conductivities, is can be even assumed that $\rho_i^{th}$ is constant with a good accuracy, as is evident, for example, from reports by Wilson et al. [86] and Vera-Londono et al. [83], where authors implemented, what they called, cross-point method for calibration of thermal contact resistance and probe-sample contact radius. The method implies that the thermal boundary resistance is constant for a set of calibration samples with thermal conductivities below ca. 1 W K⁻¹ m⁻¹. From Equation (54), larger thermal conductivities require smaller boundary resistances and/or larger contact sizes. Still, very meaningful results can be obtained by SThM for samples with a much higher thermal conductivity after an appropriate sample preparation to achieve a small variation of $\rho_i^{th}$ across the surface, such as for doped Ge samples in a work by Spièce et al. [36] ($k_{Ge} \approx$ 60 W K⁻¹ m⁻¹ at room temperature). However, uncertainty in the determination of the absolute value of the sample thermal conductivity can be very large in this case because of a much larger value of the boundary resistance (and it was not reported by Spièce et al. for the doped Ge sample). On the other hand, measurements of samples consisting of different materials, like composite ceramics, by Alikin et al. in Ref. [73] lead to a conclusion that the interfacial thermal resistances, $\rho_i^{th}$, of different materials in one and the same sample are different, and this difference is dominating in the material contrast observed in SThM images with material thermal conductivities, $k$, between 4 W K$^{-1}$ m$^{-1}$ and 30 W K$^{-1}$ m$^{-1}$. In should be noted that Vera-Londono et al., Spièce et al., and Alikin et al. used the same type of microfabricated resistive probes.

## 3. Conclusions

We have developed a calibration protocol for active SThM probes for implementation in in-vacuum measurements and mapping of sample thermal conductivity. The calibration involves measurements with a probe out-of-contact and in contact with two calibration samples. Requirements to calibration samples have been described and examples of samples suitable for the calibration have been identified in published literature. The calibration procedure yields a small set of parameters and simple mathematical expressions, which can be used to quantify thermal resistance of unknown samples as well as to characterize active-mode SThM probes of any type and at any measurement frequency. Due to this universality, the proposed calibration technique can become a standard in SThM, both for measurements with unknown samples and for characterization of probes, significantly simplifying the SThM task from the user perspective.

We have illustrated how the parameters obtained as a result of probe calibration can be used to access and optimize probe sensitivity for specific applications as well as to guide probe design. The derived relations and data about probes available on the market allow making the conclusion that by altering probe thermal resistance only, the probe sensitivity can be optimized relatively insignificantly for the SThM practice. With the probe-sample boundary thermal resistance being the main factor fundamentally limiting sensitivity, the main obstacle in measuring samples with thermal conductivities above 1 W K$^{-1}$ m$^{-1}$ is a large variations and uncertainty of the boundary thermal resistance compared to the spreading thermal resistance of a sample. The sensitivity can be increased by increasing sensitivity of probe temperature detection and increasing the maximal working temperature of a probe.

As one more concluding remark, it should also be noted that the calibration parameters cannot be a substitute for comprehensive information required for probe design and all-aspect performance metrics. A full assessment of heated cantilevers requires an understanding of the cantilever mechanical, electrical, and thermal behavior with a set of corresponding metrics,[61, 87] which remain outside of the calibration procedure developed in this work.

**Experimental section**

*FEA model details:* Numerical modeling of the probe was carried out with the use of the Joule Heating module of the COMSOL Multiphysics v.5.3a finite elements analysis package (COMSOL, AB). The dimensions and thickness of the KNT probe cantilever and the electrical leads were taken from manufacturer's specifications. The lateral dimensions of the Pd sensor strip were determined from scanning electron microscopy images in-house. The material properties were a subject of the FEA model calibration process, which is described in detail in Ref. [73]. The probe-prototype for the FEA models in this work was from the same batch as used in Ref. [73], and all the material parameters are the same as in Ref. [73] except for a somewhat higher thermal conductivity of the cantilever material (SiN$_x$), $k_{SiN}$ = 3.2 W/(m·K), and a lower TCR of Pd $\alpha_{Pd} \approx 1.6 \times 10^{-3}$ 1/K at RT = 300 K. (TCR value of bulk pure Pd is $\alpha_{Pd} \approx 3.5 \times 10^{-3}$ 1/K at 300 K.[88]) The models of this work included only the cantilever part of the KNT probe because the cantilever temperature at the cantilever base (point of cantilever attachment to the probe chip) changes negligibly during heating of the probe's Pd strip. Boundary conditions for the heat transfer problem, except those explicitly indicated in Figure 3, were thermal insulation. The numerical area of the face at the probe apex in the model is one-half of the area corresponding to a contact with a radius of 50 nm. With no sample present, the boundary conditions on this face is thermal insulation. For convenience of result post-processing, the thermal resistance of the interface between the "thermal resistor" domain and the probe apex was set to zero. In this case, the temperature at the interface between the domain and probe is a continuous function that does not experience an artificial jump, which is observed when the interfacial resistance in the model has a finite value. As in a real SThM experiment, in the numerical experiments, an electric current is injected into the gold current lead of the probe through the terminal face of the lead indicated in Figure 3a as "$I_{probe}$". Probe response is calculated as the voltage appearing on the terminal face, where the probe current is injected to monitor the resistance of the sensor strip and hence the sensor temperature. We performed modeling for states with only DC or only AC currents, without any DC offsets in the case of AC. The modeling describes only diffuse heat transfer in the sample and tip apex, without any ballistic components.

**Supporting Information**

Supporting Information is attached as well as available from the Wiley Online Library or from the author.

**Acknowledgments**

This work was supported by the author's individual support by the 2021.03599.CEECIND/CP1659/CT0016 contract (DOI:

10.54499/2021.03599.CEECIND/CP1659/CT0016) through national funds provided by FCT – Fundação para a Ciência e a Tecnologia. The work was developed within the scope of the project CICECO-Aveiro Institute of Materials, UIDB/50011/2020 (DOI: 10.54499/UIDB/50011/2020), UIDP/50011/2020 (DOI: 10.54499/UIDP/50011/2020) & LA/P/0006/2020 (DOI: 10.54499/LA/P/0006/2020), financed by national funds through the FCT/MCTES (PIDDAC).

**Conflict of Interest**

**Data Availability Statement**

The data that support the findings of this study are available from the author upon reasonable request.

**Keywords**

# Supporting Information

## A general method for calibration of active scanning thermal probes

Alexander Tselev

*CICECO – Aveiro Institute of Materials and Department of Physics,*

*University of Aveiro, 3810-193 Aveiro, Portugal*

E-mail: atselev@ua.pt

**1. Alternative port 2 in the network representation of a probe**

We can choose port 2 with terminals denoted as 3' and 4' near the input terminal 2 in the schematic in Figure 1c of the main text. This will not affect the results for port 1. Furthermore, 3, 4, 3', and 4' are connected to the same RT lead (that is, environment). With the redefined port 2, the schematic of Figure 1c in the main text remains the same; the values of matrix elements, however, become generally different. For the 3'-4' port, it is obvious that $Q' = Q$ in Equation (4) of the main text, and, hence:

$$-t_{12} = t_{11}. \qquad (1)$$

The elimination of matrix elements from equations can be performed so that $Z_x$ is expressed through $t_{12}$, which results in:

$$Z_x = \frac{a_{12} + l_{12} - t_{12}}{a_{22}(l_{12} - t_{12})} \qquad (2)$$

We also note that:

$$l_{12} = t_{12,\text{out}} = -t_{11,\text{out}} \qquad (3)$$

and we obtain essentially the same set of equations as Equations (16) of the main text:

$$\begin{cases} \beta Z_{c1} = \dfrac{a^v_{12} - t^v_{11,\text{out}} + t^v_{11,c1}}{a^v_{12}(-t^v_{11,\text{out}} + t^v_{11,c1})}, \\ \beta Z_{c2} = \dfrac{a^v_{12} - t^v_{11,\text{out}} + t^v_{11,c2}}{a^v_{12}(-t^v_{11,\text{out}} + t^v_{11,c2})} \end{cases} \qquad (4)$$

## 2. Note on calculation of Joule heat in the linearized COMSOL models

In COMSOL, the Joule heat density (volumetric electric loss density), which is $Q_{rh}$ in the COMSOL's notation, is calculated with the use of the expression:

$$Q_{rh} = \mathbf{j} \cdot \mathbf{E} \tag{5}$$

for stationary calculations and:

$$Q_{rh} = \frac{1}{2}\mathbf{j} \cdot \mathbf{E}^* \tag{6}$$

for frequency-domain perturbation calculations, where $\mathbf{E}$ is the electric field, $\mathbf{j}$ is the electric current density, and the star denotes the complex conjugation. With a material following the Ohm's law, $Q_{rh} = \mathbf{j} \cdot \mathbf{j}^*/\sigma$, where $\sigma$ is the local electrical conductivity. With an AC current as a perturbation, $\sigma$ is assumed to be constant over time in the software. Therefore, in our modeling approach, the conductivity variations are calculated as a response to temperature oscillations caused by distributed heat sources, with the heat sources being the perturbation.

As mentioned in the main text, to calculate the Joule heat sources distribution, we "preheat" the model first with a constant current of the value $I'_{DC} = I_0/\sqrt{2}$ and use the calculated state as a linearization point, at which the AC current of the amplitude $I'_{AC} = I_0/\sqrt{2}$ is applied as a perturbation to calculate the distribution of the Joule heat generation in the probe, $Q_{rh} = \mathbf{j} \cdot \mathbf{j}^*/\sigma$, with $\sigma$ distribution calculated for the "preheated" probe. And, as the last step, the calculated distribution of the Joule heat generation in the probe, $Q_{rh}$, is applied as a perturbation, again to the "preheated" probe, to calculate the resulting conductivity variations. After that, the linearized solution is still not exact, because it does not capture accurately variations of the heat generation due to variations of conductivity *during* the current oscillation period, but the solutions are accurate enough for the calibration verification.

The rationale behind this choice of $I'_{DC}$ and $I'_{AC}$ values for the current in the model is as follows. With no DC component of the current, the local Joule heat density is proportional to:

$$Q_{rh} \sim [\tilde{j}\cos(\omega t)]^2 = \frac{\tilde{j}^2}{2} + \frac{\tilde{j}^2}{2}\cos(2\omega t), \tag{7}$$

where $\tilde{j}$ is the AC current density amplitude. The term oscillating at frequency $2\omega$ in Equation (7) is the goal of the calculation at the second step. At the same time, with a non-zero DC current density $j_{00}$ and an AC current density amplitude $\tilde{j}$ at the linearization point, the Joule heat is proportional to:

$$Q_{rh} \sim [j_{00} + \tilde{j}\cos(\omega t)]^2 = j_{00}^2 + 2j_{00}\tilde{j}\cos(\omega t) + \tilde{j}^2\cos^2(\omega t)$$
$$= j_{00}^2 + 2j_{00}\tilde{j}\cos(\omega t) + \frac{\tilde{j}^2}{2} + \frac{\tilde{j}^2}{2}\cos(2\omega t) \quad (8)$$

In such a case, COMSOL automatically calculates the term oscillating at frequency $\omega$ as the linear response, and its amplitude can be used to calculate the amplitude of the oscillating term in Equation (7). It should be taken into account, however, that the COMSOL's expression for $Q_{rh}$ for the frequency-domain perturbation setup, Equation (6), has factor 1/2. That is, the calculated amplitude of the $Q_{rh}$ term at frequency $\omega$ will be equal to $j_{00}\tilde{j}/\sigma$. It is clear that to match this amplitude with the amplitude of the oscillating term of Equation (7), the DC and amplitude of AC through the probe should be both set equal to $I_0/\sqrt{2}$ in the model.

Alternatively, to calculate the linearization point, the model can be "preheated" by the stationary heat sources calculated at the first step (with the DC current equal to $I_0/\sqrt{2}$) while later, when calculating perturbations, the DC current at the linearization point is set to zero. In this case, the linear-response $Q_{rh}$ calculated by COMSOL at the second step will be zero; however, the software calculates and gives access to the fourth term in the last part of Equation (8), with the first two terms being equal to zero. The fourth term will be calculated correctly with the AC current amplitude in the model equal to $I_0$ and with factor 1/2 removed from the expression for $Q_{rh}$, Equation (6), in COMSOL. Then, factor 1/2 should be manually removed from the default COMSOL's setup. Both the approaches to perturbation calculations gave nearly indistinguishable results.

## 3. Maps of temperature distributions

*Frequency Domain* solver in COMSOL easily allows simulating and visualizing time-harmonic oscillations of various quantities in the models. Figure S1a shows a map of zero-frequency (time-averaged) component of the temperature distributions in the tip part of the standard KNT probe. This component is the same for all non-zero probe current frequencies and should be used with animated maps provided as GIF-files separately. Figures S1b and c show distributions of temperature oscillation amplitudes for frequencies 100 Hz and 100 kHz, respectively. The simulations were performed for the probe driven by an AC current with an amplitude of 1 mA and zero DC component.

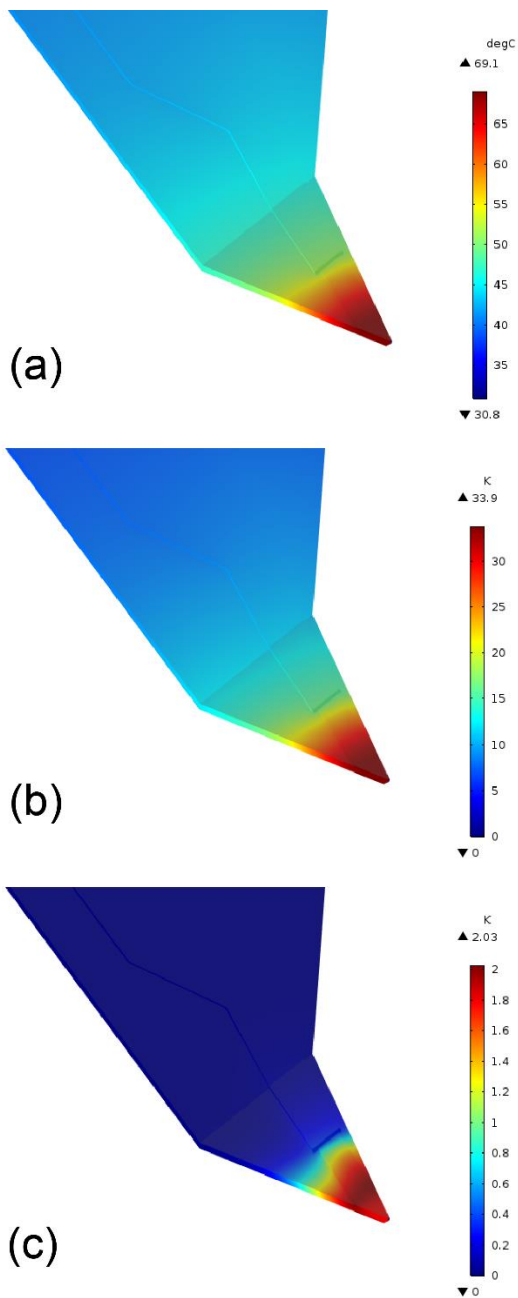

**Figure S1**. Simulated temperature distributions in the tip part of the FEA model of the standard KNT probe. The simulations are for the probe driven by an AC current with an amplitude of 1 mA and zero DC component. (a) Zero-frequency component, which is the same for all non-zero probe current frequencies. (b) Temperature amplitude distribution at a frequency of 100 Hz. (c) Temperature amplitude distribution at a frequency of 100 kHz.

## 4. Animations of simulated temperature oscillations

Separately, as a part of this Supporting Information, we provide GIF-files with animations of temperature oscillations in the tip part of the standard KNT probe calculated with the FEA models for several probe current frequencies. The simulations are for a probe driven by an AC current with an amplitude of 1 mA and zero DC component. Table S1 lists file names and corresponding probe current frequencies.

**Table S1.** Probe current frequencies corresponding to GIF-files with animated maps of temperature oscillations.

| File name | Probe current frequency (Hz) |
|---|---|
| Temperature_Variations_f1.gif | 1 |
| Temperature_Variations_f100.gif | 100 |
| Temperature_Variations_f1k.gif | 1,000 |
| Temperature_Variations_f10k.gif | 10,000 |
| Temperature_Variations_f100k.gif | 100,000 |